# Contribution of the kinetics of G proteins dissociation to the characteristic modifications of N-type calcium channel activity

**Norbert Weiss[1], Christophe Arnoult[1], Anne Feltz[2] & Michel De Waard[1*]**

[1] *Inserm U607, Laboratoire Canaux Calciques, Fonctions et Pathologies, CEA, Grenoble 38054, France ; Université Joseph Fourier, Grenoble 38054, France*
[2] *Laboratoire de Neurobiologie, UMR 8544, Ecole Normale Supérieure, Paris 75230, France*

Number of pages: 24; Number of figures: 7; Number of tables: 0

**Corresponding author:**

Dr. Michel De Waard
Inserm U607, CEA, 17 Rue des Martyrs, Bâtiment C3, 38054 Grenoble Cedex 09, France.
Tel. (33) 4 38 78 68 13
Fax (33) 4 38 78 50 41
Email: mdewaard@cea.fr

**Abstract**

Direct G protein inhibition of N-type calcium channels is recognized by characteristic biophysical modifications. In this study, we quantify and simulate the importance of G protein dissociation on the phenotype of G protein-regulated whole-cell currents. Based on the observation that the voltage-dependence of the time constant of recovery from G protein inhibition is correlated with the voltage-dependence of channel opening, we depict all G protein effects by a simple kinetic scheme. All landmark modifications in calcium currents, except inhibition, can be successfully described using three simple biophysical parameters (extent of block, extent of recovery, and time constant of recovery). Modifications of these parameters by auxiliary β subunits are at the origin of differences in N-type channel regulation by G proteins. The simulation data illustrate that channel reluctance can occur as the result of an experimental bias linked to the variable extent of G protein dissociation when peak currents are measured at various membrane potentials. To produce alterations in channel kinetics, the two most important parameters are **the extents of initial block and recovery**. These data emphasize the contribution of the degree and kinetics of G protein dissociation in the modification of N-type currents.

# 1. Introduction

N-type voltage-dependent calcium channels are strongly regulated by G protein coupled receptors (GPCR) (Dunlap and Fischbach, 1981). The $Ca_v2.2$ pore-forming subunit is a target for direct G protein inhibition (Bourinet et al., 1996). The regulation occurs through binding of the $G_{\beta\gamma}$ dimer (Herlitze et al., 1996; Ikeda, 1996) on various $Ca_v$ structural elements (De Waard et al., 1997; Qin et al., 1997; Zamponi et al., 1997). Various biophysical modifications are used for the identification of direct G protein regulation including: i) current inhibition (Boland and Bean, 1993), ii) slowing of activation kinetics (Marchetti et al., 1986), iii) depolarizing shift of the voltage-dependence of activation (Bean, 1989), iv) current facilitation following strong prepulse depolarization (Ikeda, 1991), and v) slowing of inactivation kinetics (Zamponi, 2001). At the single channel level, activated G proteins induce an increase in the first latency to channel opening and in the occurrence of sweeps without openings (Carabelli et al., 1996; Patil et al., 1996). Reduced prevalence of high opening probability gating modes has also been reported without delay in first latency opening (Delcour and Tsien, 1993). Finally, in order for the channel to recuperate full activity, $G_{\beta\gamma}$ dimer must first dissociate from the channel. This occurs during channel opening following membrane depolarization (Patil et al., 1996). However, the importance of the $G_{\beta\gamma}$ dissociation in the phenotype of N-type channels under G protein regulation requires further description. Notably, the role of G protein dissociation in producing kinetic modifications has not been simulated. Also, the contribution of G protein dissociation to the concept of channel reluctance has not been evaluated.

Here, we analyze how $G_{\beta\gamma}$ dissociation contributes to each biophysical effect. Experimental evidence is provided that the time constant of $G_{\beta\gamma}$ dissociation follows the voltage-dependence of channel opening. With a simulation approach based on three biophysical parameters of the time-dependent recovery from G protein inhibition, all landmark modifications can be described. These data suggest that current inhibition is the sole "On" effect of G proteins regulation, whereas all other effects are due to a time-dependent dissociation of $G_{\beta\gamma}$ dimer from the channel and thus are "Off" effects. These findings simplify the interpretation of calcium channel regulation by G proteins and emphasize the importance of the kinetics of G protein unbinding from the channel.

## 2. Materials and methods

### 2.1. Materials

The cDNAs used in this study were rabbit $Ca_v2.2$ (accession number D14157), rat $\beta_{2a}$ (M80545), rat $\beta_3$ (M88751) and rat $\mu$-opioid receptor (rMOR, kindly provided by Dr P. Charnet). D-Ala[2],Me-Phe[4],glycinol[5])-Enkephalin (DAMGO) was from Bachem (Bubendorf, Germany).

### 2.2. Transient expression in Xenopus oocytes

Stage V and VI oocytes were surgically removed from anesthetized adult *Xenopus laevis* and treated for 2-3 hrs with 2 mg/ml collagenase type 1A (Sigma). Cytoplasmic injection of cells was performed with 46 nl of cRNA mixture ($Ca_v2.2$ and rMOR at 0.3 $\mu g/\mu l$ $\pm$ 0.1 $\mu g/\mu l$ of $\beta_{2a}$ or $\beta_3$) *in vitro* transcribed using the SP6 or T7 mMessage mMachine Kit (Ambion, Cambridgeshire, UK). Cells were incubated at 19°C in defined nutriment oocyte medium as described (Eppig and Dumont, 1976).

### 2.3. Electrophysiological recording

After 2-4 days incubation, macroscopic currents were recorded at room temperature using Digidata 1322A and GeneClamp 500B amplifier (Axon Instruments, Union City, CA) by the two-electrode voltage-clamp technique in a bathing medium containing (in mM): $Ba(OH)_2$ 40, NaOH 50, KCl 3, HEPES 10, niflumic acid 0.5, pH 7.4 with methanesulfonic acid. Niflumic acid was used since it behaved as a voltage-independent blocker of calcium-activated chloride currents (Qu and Hartzell, 2001). Electrodes filled with (in mM): KCl 140, EGTA 10 and HEPES 10 (pH 7.2) had resistances between 0.5 and 1 M$\Omega$. Acquisition and analyses were performed using the pClamp 8 software (Axon Instruments). Recording were filtered at 2 kHz. Leak subtraction was performed on-line by a P/4 protocol. The holding potential was -90 mV throughout. For prepulse facilitation experiments, a 20 ms interpulse to -90 mV was used. DAMGO was applied at 10 $\mu M$ by superfusion of the cells at 1 ml/min. All recordings were performed within 1 min after DAMGO reached its maximum effects. By this approach, we greatly minimized voltage-independent G protein regulation that took place between 5-10 min after DAMGO application (data not shown). Only the voltage-dependent component was studied by prepulse applications.

## 2.4. Data analysis

Only cells that lacked tonic G protein inhibition and showing current densities less than 2 $\mu A/\mu F$ were included in these analyses. Furthermore, we used cells that possessed similar current densities whether expressing $Ca_v2.2$ alone or in combination with $\beta_3$ subunit. Current-voltage relationships (*I/V*) were fitted with a modified Boltzmann equation $I_{(V)} = G_{max} \times (V-E))/(1+\exp(-(V-V_{1/2})/k))$ where $G_{max}$ is the maximal conductance, E the inversion potential of the $Ba^{2+}$ current, $V_{1/2}$ the half-activation potential and k the slope factor. Prepulse-induced relief of G protein inhibition was obtained by normalizing DAMGO-inhibited currents to control currents in order to eliminate prepulse-induced inactivation, such that: $R_{I(t)} = 100 - [(I_{C(t)} - I_{DAMGO(t)}) / I_{C(t)}] / [(I_{C(t0)} - I_{DAMGO(t0)} / I_{C(t0)}]$ where $R_{I(t)}$ represents the percentage of inhibition relief by a prepulse application of variable amplitude and duration, $I_{C(t)}$, control current, $I_{DAMGO(t)}$, DAMGO-inhibited current, t the duration of the prepulse, and t0, the start of depolarization. All data are presented as mean $\pm$ SEM for *n* number observations and statistical significance (*p*) was calculated using unpaired Student's *t*-test.

## 3. Results

### 3.1. The time constant of $Ca_v2.2$ current recovery from G protein inhibition correlates with channel opening

Parameters for $Ca_v2.2$ currents recovery from G protein inhibition were determined by comparing current amplitudes during a 500 ms test pulse at 10 mV in control and DAMGO condition after application of depolarizing prepulses of variable amplitude and duration (Fig. 1). In one representative example with prepulses at 60 mV, increasing prepulse duration increases $Ca_v2.2$ currents facilitation (Fig. 1A). The prepulse-induced relief from G protein inhibition ($R_I$) was plotted as a function of prepulse voltage and duration indicating a maximal recovery from DAMGO inhibition of 45.4 ± 8.8% (n=7, prepulse of 100 mV) (Fig. 1B). Time constants of recovery were plotted as a function of prepulse potentials (Fig. 1C). The fastest time constant of recovery is observed for prepulse potential values equal or above 60 mV (70 ± 17 ms, n=7, prepulse potential of 80 mV), whereas the slowest time constant is observed at -20 mV (1101 ± 447 ms, n=7), with a half maximal current recovery time constant prepulse potential value ($PP_{1/2}$) of 10.4 ± 1.1 mV (n=56). The facilitated current being linked to channels no longer under G protein inhibition, we investigated the voltage-dependence of control channel conductance which reflects both the recruitment of channels and the increased opening probability (Fig. 1D). The half-maximal increase in conductance is 8.7 ± 1.2 mV (n=19), value which is very closely related to the estimated $PP_{1/2}$ value of 10.4 mV. Interestingly, the time constant of current recovery was well correlated to the relative conductance value of the channel (Fig. 1E, linear regression coefficient of 0.998). These data confirm the proposal that it is channel opening, and not voltage, which represents the motor of G protein dissociation (Patil et al., 1996). We further extend these initial observations by linking the kinetics of G protein dissociation to the relative conductance of the channels. As we shall see later, this observation is crucial to readdress the question of channel reluctance (Fig. 5). With such a correlation, one would also predict that a shift in the voltage-dependence of channel opening should also induce a shift in the voltage-dependence of the time constant of current recovery from G protein inhibition. **Auxiliary β subunits are known to produce such shifts in the voltage-dependence of channel activation. An endogenous β subunit (similar to $β_3$) has been described at the mRNA level in oocytes (Tareilus et al, 1997). However, it appears to be at levels too low to be detected at the protein level as assessed by metabolic labelling or Western blot experiments (data not shown). The levels were** obviously also too low to affect the biophysical parameters of Ca$_v$2.2. To assess the effects of **b$_3$** subunit, it was thus expressed exogenously.

*3.2. The **b$_3$** subunit shifts the voltage dependence of the time constant of Ca$_v$2.2 current recovery from G protein inhibition*

Auxiliary β subunits are known to displace the voltage dependence of N-type channel opening (Stea et al., 1993). We determined the effect of β$_3$ subunit on the time constant of Ca$_v$2.2 current recovery from G protein inhibition. A representative example of the effect of prepulse duration increase at 50 mV is shown for Ca$_v$2.2 / β$_3$ currents (Fig. 2A). The prepulse-induced relief of G protein inhibition was quantified by measuring R$_I$ values (Fig. 2B). The maximal relief of DAMGO inhibition, observed for Ca$_v$2.2 / β$_3$, was 48.1 ± 11.5% (n=6, prepulse of 100 mV), which is comparable to the relief observed for Ca$_v$2.2 alone cells. Moreover, the faster time constant of recovery is observed for prepulse potential values equal or above 40-60 mV (17 ± 2 ms at a prepulse potential of 100 mV, n=4), **a time constant which is 4.1 faster than the one observed for Ca$_v$2.2 channels alone (Fig. 2C).** The ratios of recovery time constants for Ca$_v$2.2 and Ca$_v$2.2 / β$_3$ channels as a function of prepulse potential illustrate that **b$_3$ induces a faster recovery from G protein inhibition with a maximal effect of 6.1-fold at 22 mV (Fig. 2C inset). Similar acceleration of recovery was observed previously** (Roche and Treistman, 1998). However, it is the first time that it is shown that this effect of β subunit on the time constant is not stable over the range of potentials known to activate N-type channels. The half maximal current recovery time constant prepulse potential value (PP$_{1/2}$) for Ca$_v$2.2 / β$_3$ channels is -2.4 ± 1.0 mV (n=22) which is shifted by -12.8 mV compared to Ca$_v$2.2 channels (Fig. 2C). This shift is reminiscent of the shift produced by β subunits on the voltage dependence of Ca$_v$2.2 channel activation (Stea et al., 1993). Indeed, half-maximal increase in conductance of Ca$_v$2.2 / β$_3$ channels occurs at -1.5 ± 1.5 mV (n=10, Fig. 2D) which is thus shifted by -10.2 mV compared to the half-maximal conductance of Ca$_v$2.2 channels. The time constant of current recovery was indeed correlated to the relative conductance value of the channel (Fig. 2E, r = 0.98).

*3.3. The **b$_3$** subunit does neither antagonize nor promote G protein inhibition*

One main problem in examining the effect of β subunits on G protein regulation is that it also affects channel expression levels. With higher expression levels, the bioavailability of endogenous G proteins becomes limiting (data not shown). Here, we deliberately avoided this pitfall by recording from cells possessing low (maximal bioavailability of G proteins) and similar currents densities ($1.41 \pm 0.33$ µA/µF for $Ca_v2.2$ and $1.39 \pm 0.50$ µA/µF for $Ca_v2.2$ / $\beta_3$). In spite of this precaution, the inhibition at the peak of the currents is less pronounced in the presence of $\beta_3$ subunit (mean inhibition of $30.4 \pm 4.9\%$ at 30 mV (n=49) with $\beta_3$ *vs* $54.3 \pm 5.5\%$ (n=109) without $\beta_3$) (Fig. 3A,B). Such a reduction has often been interpreted as due to an antagonistic effect of $\beta$ subunit (Bourinet et al., 1996). It is known that a faster recovery from inhibition (5.6-fold at 30 mV with $\beta_3$ subunit (Fig. 2C, inset)) may strongly diminish the DAMGO inhibition when measured at the peak current ((Roche and Treistman, 1998); and present data). In order to evaluate whether $\beta_3$ subunit influences the G protein inhibition in our conditions of similar expression levels, we followed the percentage of DAMGO inhibition as a function of depolarization time at 30 mV both for $Ca_v2.2$ and $Ca_v2.2$ / $\beta_3$ channels (Fig. 3C). Extrapolating the curves to the start of depolarization, provides an indication of what the real DAMGO inhibition should be before the start of the recovery process. It appears that the DAMGO inhibition at the start of depolarization for $Ca_v2.2$ and $Ca_v2.2$ / $\beta_3$ channels is not statistically different ($66.2 \pm 5.4\%$ *vs* $67.8 \pm 5.7\%$ respectively) (Fig. 3D). DAMGO inhibition at the peak of the current is thus considerably under-evaluated by 11.9% and 37.4% for $Ca_v2.2$ and $Ca_v2.2$ / $\beta_3$ channels respectively. These data extend earlier observations by demonstrating that $\beta_3$ subunit is neither an antagonist nor an agonist of G protein current inhibition. However, faster recovery at lower voltages significantly boosts G protein dissociation.

*3.4. Simulation of current kinetics with parameters affecting the recovery from DAMGO inhibition*

Recovery from G protein inhibition takes place during channel opening and this recovery is known to affect the activation kinetics of the current (Elmslie and Jones, 1994). However, little has been done so far to simulate how G protein dissociation may influence channel activation kinetics and inactivation kinetics alike. We therefore simulated the type of modifications introduced in the current by the progressive recovery from G protein inhibition during membrane depolarization. In our simulation studies, we made one reasonable assumption: blocked channels are non permeable channels and should contribute to a control-like current after G protein dissociation. This assumption is supported by the single channel data from the group of Dr Yue (Patil et al., 1996). Hence, non blocked currents at the start of the depolarization are supported by non-regulated channels. The concept that channels may open with low probability in a reluctant mode while G proteins are presumably bound onto the channel (Lee and Elmslie, 2000) is not introduced in our simulation work for reasons that will be discussed latter on. Thus, currents recorded under DAMGO inhibition should represent the sum of currents flowing through two different populations of channels: current from Non Regulated channels ($I_{NR}$), which possess similar properties than control channels, and current from channels undergoing a Progressive Relief from G protein inhibition ($I_{PR}$). Thus:

$$I_{DAMGO} = I_{NR} + I_{PR} \tag{1}$$

where $I_{NR}$ represents a fraction of the control current such that:

$$I_{NR} = I_{Control} \times ((100\text{-}DI)/100) \tag{2}$$

where DI is the percentage of DAMGO inhibition (DI = 54.3% for $Ca_v2.2$ alone and 30.4% for $Ca_v2.2$ / $\beta_3$ channels at 30 mV; both values being under-estimated as we have quantified above).

We next described $I_{PR}$ as the blocked fraction of $I_{Control}$

$$I_{Blocked} = I_{Control} \times (DI/100) \tag{3}$$

that would progressively recover from G protein inhibition with the following time course:

$$\text{Time course of recovery from block} = 1\text{-}\exp(\text{-}(1/\tau)\times t) \tag{4}$$

where t is the time after depolarization and $\tau$ is the time constant of inhibition relief.

Since only a fraction $R_I$ of the blocked current really recovers, then:

$$I_{PR} = I_{Blocked} \times ((1\text{-}\exp(\text{-}(1/\tau)\times t))\times(R_I/100)) \tag{5}$$

We thus can propose the equation 6 with:

$$I_{DAMGO} = [I_{Control}\times(100\text{-}DI)/100] + [I_{Control}\times(DI/100)\times(1\text{-}\exp(\text{-}(1/\tau)\times t))\times(R_I/100)] \tag{6}$$

This equation has thus three different parameters that can be modulated to describe $I_{DAMGO}$: DI, $\tau$ and $R_I$. In Fig. 4A-C, we illustrate how varying each of these three parameters influences the kinetics of a representative $Ca_v2.2 / \beta_3$ currents at 30 mV. When one parameter was varied, the two others were fixed at their mean experimental values (i.e., $\tau = 27$ ms (Fig. 2C), $R_I = 48\%$ (Fig. 2B) and DI = 31% (Fig. 3B)). In Fig. 4Aa, DI was fixed at 30, 50 or 70%. As expected with such a time constant, 48% of the blocked channels would recover from inhibition at the end of a 500 ms pulse. We also noticed that the greater the fraction of inhibited channels, the greater the shift of the time to peak (by 13.3 ms for 30% inhibition, 26.5 ms for 50% inhibition, and 37.8 ms for 70% inhibition in this representative example). This tendency was illustrated by plotting the simulated shift of time to peak as a function of DAMGO inhibition (Fig. 4Ab). With our representative control $Ca_v2.2 / \beta_3$ current trace at 30 mV, the maximal shift of time to peak that can theoretically be predicted is 54.9 ms considering 100% of inhibition by DAMGO (fixed parameters of $\tau = 27$ ms and $R_I = 48\%$). In Fig. 4Ba, we fixed $R_I$ at 30, 50 or 70%. Interestingly, in this configuration, this parameter has a smaller impact on the shift of the time to peak of the current (shift of 13.3 ms for $R_I = 30\%$, 16.4 ms for $R_I = 50\%$, and 24.5 ms for $R_I = 70\%$). Plotting the theoretical shift of time to peak as a function of $R_I$ values illustrates that a maximal shift of 26.5 ms can be expected (Fig. 4Bb; fixed parameters $\tau = 27$ ms and DI = 31%). In Fig. 4Ca, we fixed the time constant of recovery from inhibition at $\tau = 10$ ms, 50 ms and 250 ms (incremental 5-fold changes). We observed that this time constant has also an incidence on the time to peak of the current (shifts of 6.6 ms for $\tau = 10$ ms, 13.3 ms for $\tau = 50$ ms, and 0 ms for $\tau = 250$ ms). Again, we plotted the theoretical shift of the time to peak as a function of the time constant of inhibition relief. Fig. 4Cb illustrates a bell-shaped curve for the shift of the time to peak with an optimum at $\tau = 55.7$ ms (fixed parameters of $R_I = 48\%$ and DI = 31%). Very slow recovery time constants, above 250 ms, result in no shift at all. In contrast, such time constants affect the rate of inactivation of the channel (Fig. 4Ca). Such an effect may explain the mild G protein effects reported on inactivation kinetics (Zamponi, 2001). The rising phase of the curve illustrates that recovery $\tau$ from G protein inhibition should be slower than the activation time constant to observe a slowing of activation kinetic, whereas the descending phase illustrates that $\tau$ should be faster than inactivation. We would predict that this descending phase would not exist if the channel did not undergo inactivation. This was indeed confirmed when using a $Ca_v2.2 / \beta_{2a}$ combination of channel subunits that has very slow inactivation (data not shown). We thus conclude that the optimal conditions to observe an important shift of the time to peak should be: i) a maximal inhibition by DAMGO, ii) a maximal relief from inhibition during depolarization, iii) a time constant of relief ideally placed between the rates of channel activation and inactivation, and iv) no or little channel inactivation. The same conditions result in the maximum gain of current compared to traces that would theoretically not undergo recovery from inhibition. At this stage, it is important to notice that a slower time constant of recovery from inhibition results in less gain of current. Such a difference may produce drastic differences in current recovery between $Ca_v2.2$ and $Ca_v2.2$ / $\beta_3$ channels since the latter has faster recovery from inhibition.

The simulated DAMGO current should describe the experimental DAMGO trace regardless of the inactivation properties. To illustrate this point, we used two extreme inactivation conditions, one using a +30 mV trace from a $Ca_v2.2$ / $\beta_3$ combination (fast inactivating current), and another from a $Ca_v2.2$ / $\beta_{2a}$ combination (slow inactivating current). Fig. 4Da illustrates an experimental $Ca_v2.2$ / $\beta_3$ DAMGO current (along with the control trace) (left panel), that is compared to the simulated current without recovery from G protein inhibition (blue current trace) and to the simulated current with recovery from G protein inhibition (red trace). The parameters used for the latest trace are within the fluctuation range of observed individual values for experimental current traces at 30 mV ($R_I$ = 52% and $\tau$ = 45 ms) or coherent with the example chosen (DI = 63%). Fig. 4Db similarly illustrates a comparison between experimental (left panel) and simulated (right panel) DAMGO inhibited $Ca_v2.2$ / $\beta_{2a}$ current at 30 mV. Although the experimental value for the time constant of recovery from G protein inhibition is unknown, all the parameters used here to construct the simulated DAMGO inhibited $Ca_v2.2$ / $\beta_{2a}$ current are compatible with the observations made on $Ca_v2.2$ / $\beta_3$ channels (DI = 55%, $R_I$ = 65% and $\tau$ = 38 ms). Overall, this model, which is based on these three parameters, is able to describe real DAMGO inhibited currents regardless of the kinetics of inactivation.

*3.5. Increasing extent of recovery from G protein inhibition during increasing depolarizing steps induces an apparent reluctant state of the channel*

According to Figs. 1 and 2, faster G protein dissociations are expected for voltages that induce greater opening channel probabilities rendering the gain of current non homogenous over a range of studied potentials. Current-voltage relationships constructed under DAMGO inhibition should contain two different components: residual, non-blocked currents, with properties similar to control currents, and a variable fraction of currents that have recovered from G protein inhibition. The problem with building such current-voltage relationships is that current measures are generally performed at variable times after the start of the depolarization (peak values) and thus also with variable proportions of recovered currents. These factors have the potential to alter the shape of the current-voltage relationship. Here, we used our simulation to investigate how current recovery from G protein inhibition could affect the shape of the current-voltage relationship (Fig. 5). To illustrate this point, we investigated the voltage-dependence of recovered $Ca_v2.2$ / $\beta_3$ currents. Qualitatively similar results were obtained in the absence of $\beta_3$ subunit (data not shown). Fig. 5A illustrates an example of representative control and DAMGO-inhibited currents for a set of depolarization values (from -10 to 40 mV by 10 mV steps; left panel). **The DAMGO-inhibited currents were simulated with progressively smaller values of time constants of recovery when membrane depolarization was increased (values provided in legend of Fig. 5).** Simulated DAMGO-inhibited currents all describe accurately their real counterparts (middle panel). For a range of potential values (0-10 mV), the slow time constants of recovery (Fig. 2C) all produce a slowing of inactivation kinetics, consistent with our former conclusion (Fig. 4C). The recovered currents are also illustrated as a function of test potentials (Fig. 5A, right panel). It is clearly apparent that the maximal currents in response to recovery occur at potentials of 10 and 20 mV. Symbols illustrate the time position at which maximal DAMGO-inhibited currents were measured (see corresponding symbols in middle panel). We next constructed the current-voltage relationships in a conventional manner (i.e. at the time to peak of the current; Fig. 5B). Fig. 5B left panel illustrates the experimental control and DAMGO-inhibited currents for this representative $Ca_v2.2$ / $\beta_3$ example. Half-activation potentials were -2.8 mV (Control) and 4.1 mV (DAMGO) showing the classical depolarizing shift observed with direct G protein inhibition. Similarly, the simulated DAMGO-inhibited current has a half-activation potential of 4.4 mV (Fig. 5B middle panel). We next extracted the current-voltage relationship for currents that recovered from DAMGO inhibition (symbols in right panel of Fig. 5A). Interestingly, the current-voltage relationship of this recovered current is shifted towards depolarized values with a half-activation potential of 7.8 mV (Fig. 5B right panel). We thus conclude that the shift observed under DAMGO inhibition is largely the result of a current recovery from G protein inhibition. We term this effect "apparent reluctance" since it is not correlated to any particular state of the channel. One prediction of this $G_{\beta\gamma}$ dissociation mechanism is that the "reluctance" observed at the peak of the current should have almost disappeared if current-voltage relationships are constructed after maximal recovery from inhibition. This was indeed the case for experimental and modeled currents when measured at the end of the 500 ms pulses (Fig. 5C).

### 3.6. Fit of the G protein inhibited currents permits the extraction of the time constant of recovery from inhibition

Since DAMGO traces can be simulated by equation 6, we determined whether G protein inhibited currents could also be fit by this equation. Two of the parameters of G protein inhibition can be readily extracted from the experimental traces from expressed $Ca_v2.2$ / $b_3$ channels (Fig. 6A). At each voltage, DI values were estimated according to the method shown in Fig. 3C. As expected, since DI represents the total inhibition before depolarization, the values measured for DI were voltage-independent. The $R_I$ values were estimated at the end of the 500 ms depolarization as the percentage of recovery from G protein inhibition according to the position of the DAMGO trace with regard to the position of control current (maximal recovery) and that of the theoretical trace without recovery from inhibition (blue traces). These two measured parameters were then introduced into equation 6 in order to extract the time constant of recovery $t$ by fitting the experimental DAMGO traces at each voltage. An example of the good adequacy of the fit is provided in Fig. 6A (right panel). Average time constant of recovery were then plotted as a function of test potential (Fig. 6B). Fitting the data with a sigmoid curve provides a potential of half-maximal time constant of -2.6 ± 1.3 mV (n=10), which is not significantly different than the value obtained in Fig. 2C with the prepulse method. Similarly, a plot of the time constant of recovery from inhibition as a function of the relative channel conductance indicates a linear correlation (r=0.992). Interestingly, extracting the time constant of recovery according to this fit yields smaller values than with the prepulse method. These differences in absolute values appear to be linked to two potential problems that may occur with the prepulse method. First, the control DAMGO current (without prepulse application) is already a facilitated current since recovery from inhibition occurs at low potentials. Second, reassociation of G proteins may occur to some extent during application of the interpulse potential at -90 mV. Taken together, these two factors contribute to underestimate the amount of recovery and its kinetics.

## 4. Discussion

This study illustrates that, besides slowing of activation kinetics (Elmslie and Jones, 1994) and prepulse facilitation (Ikeda, 1991), the shift of the voltage-dependence of activation and the changes in inactivation kinetics can also be attributed to a recovery from G protein inhibition. **Figure 7 illustrates under the form of a diagram the various molecular states of the G protein regulated channel that may be at the origin of the biophysical modifications induced by G proteins. The diagram emphasizes the importance of an open- non-conducting G protein-bound state of the channel for G protein regulation. The data on the shift of the voltage-dependence of activation and on the changes in inactivation kinetics represent interesting new findings since these effects are generally attributed to G protein association ("On" effects) linked to the promotion of a channel reluctant mode, rather than to G protein dissociation ("Off" effects).** Recovery from inhibition is shown to occur with a time constant that is directly correlated with the channel opening state. Thus, the voltage-dependence of this recovery time constant is largely due to the voltage-dependence of channel opening, making $G_{\beta\gamma}$ dissociation **mechanistically voltage-independent.** The $\beta_3$ subunit shifts the voltage-dependence of this time constant of recovery towards hyperpolarized potentials and increases the fastest time constant of recovery by a factor of 4.1. The shift of the voltage-dependence is readily explained by the effect of $\beta_3$ subunit on the voltage-dependence of activation. The **acceleration of the time constant of recovery produced by b$_3$ subunit** is probably linked to the promotion of a greater channel opening probability (Wakamori et al., 1999). Looking at current traces under DAMGO application, landmark effects, other than current inhibition, can be simulated by the combination of three parameters: DI, percentage of real DAMGO inhibition, $R_I$, the extent of current recovery from inhibition, and $\tau$, the time constant of recovery. The evaluation of DI is often under-evaluated since current inhibition is usually assessed at the peak of the inhibited current. This was particularly the case in the presence of $\beta_3$ subunit which dramatically speeds the time constants of recovery at intermediate potentials, thereby leading to an apparent antagonism when measured at the peak of the current (Fig. 3A.B and see conclusions by (Roche and Treistman, 1998)). Indeed, measuring currents at their peak can severely affect the evaluation of the real extent of G protein inhibition in particular if $\tau$ is fast or if the channel has slow inactivation kinetics (like in the presence of a $\beta_{2a}$ subunit). To observe marked landmark effects, the best conditions are a strong current inhibition (DI), an important relief of inhibition ($R_I$) and a fast recovery $\tau$, comprised between the rates of channel activation and inactivation. The two latter parameters cannot be controlled experimentally, since they are intrinsic channel and G protein properties. $R_I$ is probably linked to inactivation since this process is expected to prematurely stop the recovery from inhibition. These two parameters thus appear as crucial for the bioengineering of G protein mimicking peptides with desirable N-type channel blocking properties.

*4.1. G protein unbinding is a crucial step for the biophysical changes in current properties*

It is thus clear that all landmark effects used to identify direct G protein regulation are due to G protein dissociation from the channel, with the notable exception of current inhibition. Here, the willing mode is a non-inhibited channel with unbound G protein (no different than control channel). The channel is in a fully blocked state when a G protein is bound onto it. The reluctance of opening only reflects the time course of G protein unbinding, but without any modification in the voltage-dependence of the channel. An additional degree of complexity could be introduced in our simulation data by assuming that some of the G protein bound channels can enter a low probability mode of opening (Lee and Elmslie, 2000). This degree of complexity was not introduced because we believe that the single channel activities recorded at intermediate and high potentials result from the faster G protein dissociation that occurs at these potentials. The problem with the concept of a low opening probability mode of the channel in the G protein bound state is that no proof can be provided that the channels still bind the G proteins in question. In addition, the level of consensus on this matter is rather low since contradictory reports are numerous. In particular, evidence for channel openings in a "reluctant mode" at low potential values is inexistent, most probably because G protein dissociation is very slow (Carabelli et al., 1996; Patil et al., 1996).

The question remains on what triggers G protein unbinding? The strict correlation, demonstrated here for the first time, between the voltage-dependence of G protein dissociation time constant and channel opening indicates that it is either (i) any step in the complex molecular process of opening, or (ii) opening itself, which leads to G protein dissociation. Possibility is left for G proteins to act as "gating modifiers" to inhibit channel activity as former evidence possibly indicate (Delcour and Tsien, 1993; Colecraft et al., 2000). However, G proteins may also act as "pore blockers", a hypothesis supported by the observation that G proteins alter the permeation pathway of N-type channels (Kuo and Bean, 1993). Since a 35 amino acid truncated version of $G_\beta$ protein is sufficient to confer channel inhibition, complex patterns of protein interactions do not seem necessary to provide regulation (Li et al., 2005). Nevertheless, because recovery from inhibition occurs also at potential above the reversal potential, ion influx is not the driving force for G protein dissociation.

### 4.2. The time constant of recovery from G protein inhibition can be readily extracted by fitting DAMGO traces with the simulation equation

**Since DI and $R_I$ values can be extracted directly from the comparison of experimental control and DAMGO-inhibited traces, the entire DAMGO trace can be fitted using our simulation equation thereby providing an estimate of the time constant of recovery. Using this method, the linear correlation observed between the time constant of recovery and the relative conductance is well conserved as well as the voltage dependence of the process. As discussed in the result section, differences are however noted with regard to the absolute values that should be linked to some of the artifacts introduced by the prepulse method. Globally, much faster recovery appears to exist during a depolarization suggesting that the parameters extracted from the prepulse method may have to be interpreted cautiously. In any case, it should be emphasized that the simulation method introduced herein is an alternative method for the quantitative evaluation of G protein regulation parameters.**

### 4.3. Predictions for atypical direct G protein regulations

Our data raise intriguing new scenarios. Firstly, a case in which $G_{\beta\gamma}$ would dissociate with the same rate of activation time constant of the channel. One would conclude to the lack of G protein regulation since no landmark modification should be observed (not even current inhibition) despite the presence of proven binding elements. L-type channels may belong to this category since they possess $G_{\beta\gamma}$ binding sites (Ivanina et al., 2000). This point will merit further investigation to assess whether current recovery from inhibition varies to a great extent depending on G protein and calcium channel isoforms. Secondly, a case in which the time constant of recovery from G protein inhibition is ultra-slow. Here also, all landmark effects, except current inhibition, would be absent. Prepulse facilitation would be lacking though the inhibition is direct by essence. As such, some forms of voltage-independent G protein regulation thought to be indirect may also have to be reinterpreted.

### 4.4. Concluding remarks

Providing a simple and easy to understand mechanistic frame of N-type channel regulation by G proteins, that can accommodate former models of regulation, will be useful for defining rationale structure-function studies. Considering the increased importance taken by $G_{\beta\gamma}$ dissociation in G protein regulation, it will be essential to identify the molecular determinants that drive $G_{\beta\gamma}$ off from the channel. These determinants are not voltage-dependent but are linked to the transition steps leading to channel opening.

**Acknowledgements**

We acknowledge financial support from Inserm and the French Ministry of Research and Technology. The clone coding for rat µ-opioid receptor was kindly provided by Dr. Pierre Charnet (CNRS, Montpellier, France).

**Footnotes**

The following abbreviations have been used. GPCR: G protein coupled receptors; DAMGO: D-Ala², Me-Phe[4], glycinol[5])-Enkephalin.

**Figures legends**

Fig. 1. The time constant of relief from G protein inhibition is correlated to channel opening. (A) Experimental protocol to measure prepulse facilitation as a function of prepulse duration and voltage (see *Materials and methods* for details). Representative current traces recording at 60 mV prepulse potentials are shown for $Ca_v2.2$ channels for control (left panel) and DAMGO (right panel) conditions at various prepulse durations. (B) The proportion of inhibition relief was calculated as a function of prepulse duration for many different prepulse potentials (from -20 mV to 100 mV, n=56 cells) (see *Materials and methods* for details). The data were fitted by an exponential growth according to $R_I = R_{Imax} \times (1-\exp(-(1/\tau) \times t_{PP}))$ where $\tau$ is the time constant of inhibition relief, $R_{Imax}$ the maximal proportion of relief and $t_{PP}$ the prepulse duration. (C) Time constant $\tau$ as a function of prepulse potential value (n=56 cells) with a test pulse at 10 mV. A sigmoid fit of the data yield a half-maximal **acceleration of the time** constant at $PP_{1/2} = 10.4 \pm 1.1$ mV. **The sigmoid equation used for the fit is $t = t_0 + (a/(1+\exp(-(PP-PP_{1/2})/b))$) where $a = 1038$ ms and $b = -7.9$ mV**. (D) Increase in the macroscopic conductance of $Ca_v2.2$ as a function of test potential in control condition (n=56 cells). A Boltzmann fit of the data yield a half-activation potential of $8.7 \pm 1.2$ mV (n=19 cells). (E) Correlation between the time constant of inhibition relief and the relative conductance of $Ca_v2.2$ at potential values between -20 and 60 mV. The data were fitted by a linear regression (r = 0.998).

Fig. 2. The auxiliary $\beta_3$ subunit displaces similarly the voltage-dependence of $Ca_v2.2$ activation and **the acceleration of the time** constant of recovery from G protein inhibition. (A) Experimental protocol to measure prepulse facilitation as a function of prepulse duration and voltage (see *Materials and methods* for details). Representative current traces recording at 50 mV prepulse potentials are shown for $Ca_v2.2$ / $\beta_3$ channels for control (left panel) and DAMGO (right panel) conditions at various prepulse durations. (B) Proportion of recovery from DAMGO inhibition as a function of prepulse duration for many different prepulse potentials (from 0 to 100 mV, n=22 cells). (C) Time constant $\tau$ as a function of prepulse potential value (n=22 cells). **A sigmoid fit of the data with the equation $t = t_0 + (a/(1+\exp(-(PP-PP_{1/2})/b))$) yield a half-maximal **acceleration of the** time constant at $PP_{1/2} = -2.4 \pm 1.0$ mV, $a = 641$ ms and $b = -7.7$ mV**. Inset: **modification of the time** constant of recovery by $\beta_3$ subunit. (D) Macroscopic conductance of $Ca_v2.2$ / $\beta_3$ as a function of test potential in control

condition (n=10 cells). A Boltzmann fit of the data yield a half-activation potential of -1.5 ± 1.5 mV. (E) Correlation between the time constant of recovery from inhibition and the relative conductance of $Ca_v2.2$ / $\beta_3$ channels at various potential values. The data were fitted by a linear regression (r = 0.98).

Fig. 3. Faster recovery from G protein inhibition with $\beta_3$ subunit explains $\beta$ antagonism on DAMGO inhibition. (A) Representative current traces at 30 mV for $Ca_v2.2$ (top panel) and $Ca_v2.2$ / $\beta_3$ (bottom panel) before and after 10 µM DAMGO application. (B) Average current inhibition measured at the peak of the currents at 30 mV for cells expressing $Ca_v2.2$ alone or $Ca_v2.2$ in combination with $\beta_3$. The difference in inhibition is statistically significant (p = 0.005). (C) Average DAMGO inhibitions as a function of time after the start of the 30 mV depolarization. Data were fitted with decreasing exponential. (D) Average DAMGO inhibition evaluated at the start of depolarization for $Ca_v2.2$ and $Ca_v2.2$ / $\beta_3$ channels.

Fig. 4. Simulated impact of recovery from G protein inhibition on current kinetics. (A) Effect of DI variation on current kinetics. Other parameters remained constant ($R_I$ = 48% and $\tau$ = 27 ms). *a*, Representative examples are shown on a simulated $Ca_v2.2$ / $\beta_3$ current at 30 mV. Black trace is the control trace without DAMGO; blue, simulated trace without recovery from G protein inhibition; and red, modeled DAMGO trace with current recovery. Inset: expanded traces showing the first 40 ms of the depolarization. Symbol positions indicate the time to peak of the current and, upper values, the shift of the time to peak. *b*, Effect of DI variation on the shift of the time to peak is shown. Data are issued from arbitrary DI values. (B) Effect of $R_I$ variation on current kinetics (with fixed DI = 31% and $\tau$ = 27 ms). *a*, Representative traces and *b*, effect of $R_I$ variation on the shift of the time to peak. (C) Effect of $\tau$ variation on current kinetics (with fixed DI = 31% and $R_I$ = 48%). *a*, Representative traces and *b*, effect of $\tau$ variation on the shift of the time to peak. (D) Quality of the simulation in various conditions of channel inactivation. *a*, Experimental $Ca_v2.2$ / $\beta_3$ traces at 30 mV (control and DAMGO-inhibited, left panel) and modeled traces (right panel) showing a good fit by the model (DI = 63%, $R_I$ = 52% and $\tau$ = 45 ms). *b*, Similar simulation but with a $Ca_v2.2$ / $\beta_{2a}$ trace at 30 mV (DI = 55%, $R_I$ = 65% and $\tau$ = 38 ms). The times to peak are significantly higher with slowly inactivating channels (14.1 ms of shift with $\beta_3$ and 72.2 ms with $\beta_{2a}$).

Fig. 5. Effect of current recovery from G protein inhibition on the current-voltage relationship. (A) Experimental $Ca_v2.2 / \beta_3$ current traces at various potentials (control and DAMGO inhibition, left panel), simulated traces (without (blue) and with recovery (red), middle panel) and subtracted simulated traces (red-blue=grey, right panel) illustrating the total recovered current. Symbols are positioned at the peak of the current (left and middle panel). Filled grey symbols in right panel are positioned at the time to peak of the DAMGO inhibited current traces, and not at their own peak values. **Simulated curves in red were obtained with the following parameters (DI in % / t in ms / $R_I$ in %): -10 mV (63 / 87 / 35), 0 mV (70 / 68 / 50), 10 mV (62 / 32 / 53), 20 mV (58 / 15 / 57), 30 mV (58 / 8 / 62) and 40 mV (54 / 4 / 60).** (B) Corresponding current-voltage relationships. In the right panel, we also plotted the current-voltage relationship for DAMGO inhibited currents that would not recover from G protein inhibition (blue symbol) and for recovered current (grey symbol). Data were fitted with Boltzmann equations and provide the following $V_{1/2}$ half-activation potentials (-2.8 mV for experimental control, 4.1 mV for experimental DAMGO, 4.4 mV for simulated DAMGO, 7.8 mV for recovered current and 1.3 mV for non recoverable blocked current measured at the peak of experimental DAMGO current traces). (C) Current-voltage relationships in control and experimental (left panel) or simulated (right panel) DAMGO condition for measures at the end of the depolarisation (500 ms). Half activation values are $V_{1/2}$ of -0.4 mV (control, open square), 1.7 mV respectively (experimental DAMGO, black square) and 1.2 (simulated DAMGO, red square).

**Fig. 6. Extracting the time constant of G protein dissociation from DAMGO-inhibited traces using the simulation model. (A) Left panel: representative experimental $Ca_v2.2 / b_3$ current traces at various potentials (control and DAMGO inhibition). Middle panel: experimental traces superimposed with the simulated traces without recovery (blue). DI values were measured as defined in Fig. 3C. $R_I$ values were extracted at the end of pulse by measuring the percentage of recovery as estimated by the position of the DAMGO trace between the blue simulated trace and the control experimental trace. Right panel: fit (red trace) of the DAMGO trace (black enlarged trace) by using equation $I_{DAMGO} = [I_{Control} \times (100-DI)/100] + [I_{Control} \times (DI/100) \times (1-exp(-(1/t) \times t)) \times (R_I/100)]$ in which the experimental DI and $R_I$ values were introduced as fixed parameters for each potential value and t was the variable to extract. (B) Time constant t as a function of potential value (n=10 cells). A sigmoid fit of the data with the equation $t = t_0 + (a/(1+exp(-(PP-$**

PP$_{1/2}$)/b)) yield a half-maximal speeding up of the time constant at V$_{1/2}$ = -2.6 ± 1.3 mV, a = 161 ms and b = -10.5 mV. (C) Correlation between the time constant of inhibition relief and the relative conductance of Ca$_v$2.2 / **b**$_3$ at potential values between -10 and 40 mV. The data were fitted by a linear regression (r = 0.992).

Fig. 7. Kinetic diagram illustrating the various steps involved in G protein dissociation from the channel. In this scheme, the only conducting state of the channel is the one in which G$_{bg}$ dissociation has occurred while the channel is still open. The transition from a G$_{bg}$ bound state of the channel in its closed configuration to a G$_{bg}$ unbound state in its open transition requires the passage through an additional transient G$_{bg}$ bound open- and non-conducting state of the channel. The passage through this state explains the delay in the time to peak of the current and the slowing of inactivation kinetics. Inactivation can occur in the presence of G$_{bg}$ not necessarily with the same rate constant as in the absence of G$_{bg}$. G$_{bg}$ dissociation is intrinsically voltage-independent, but the rate of dissociation becomes voltage-dependent as the result of the voltage-dependence of channel open probability.

**A**

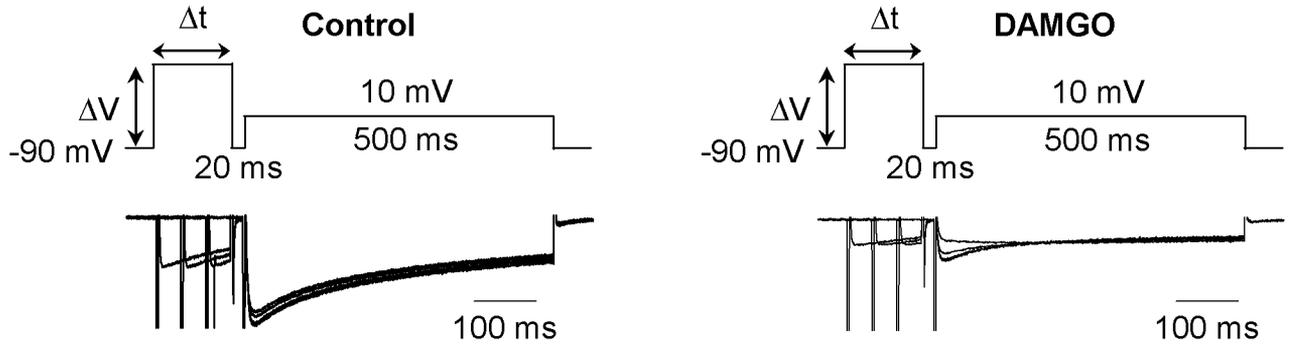

**B**

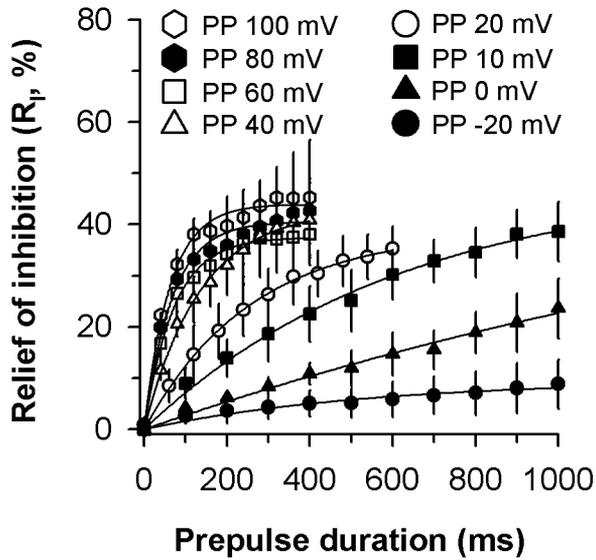

**C**

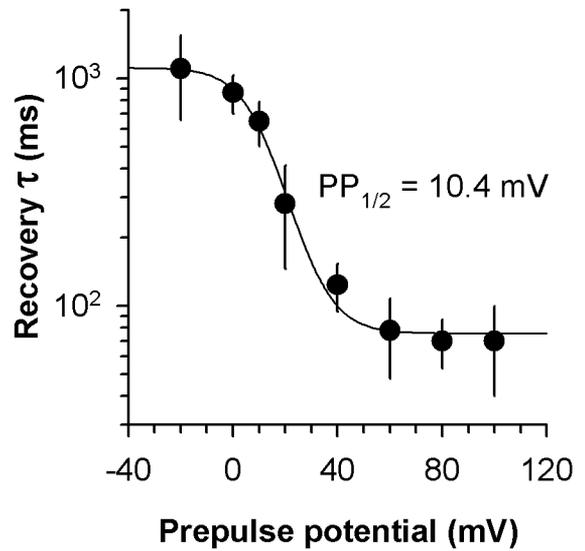

**D**

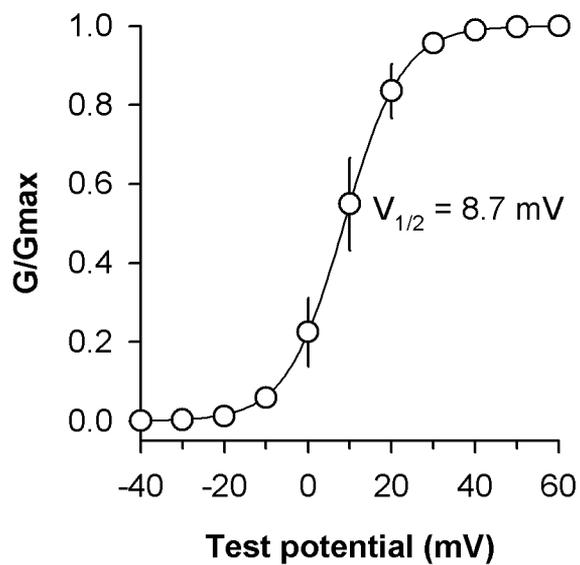

**E**

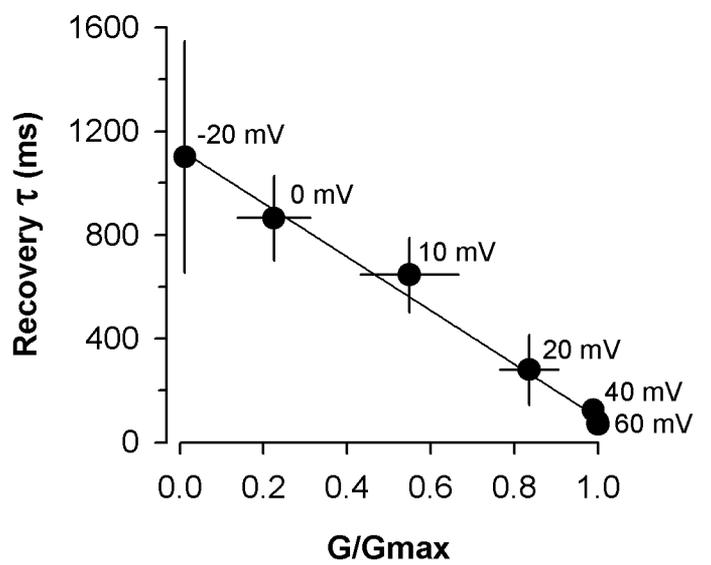

**A**

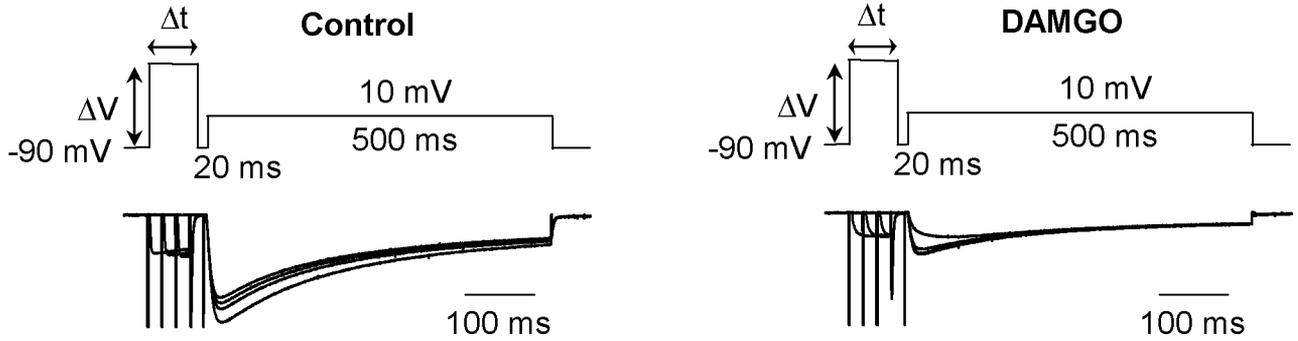

**B**

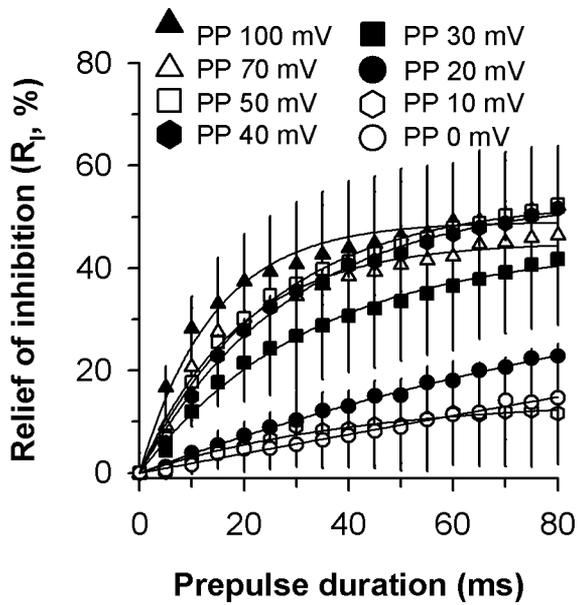

**C**

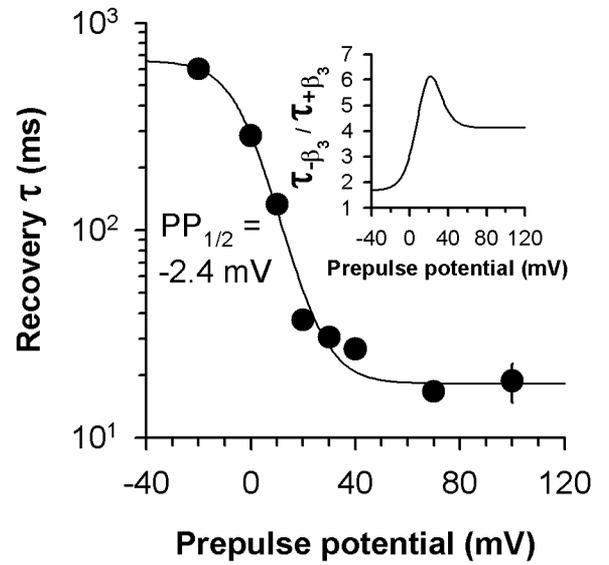

**D**

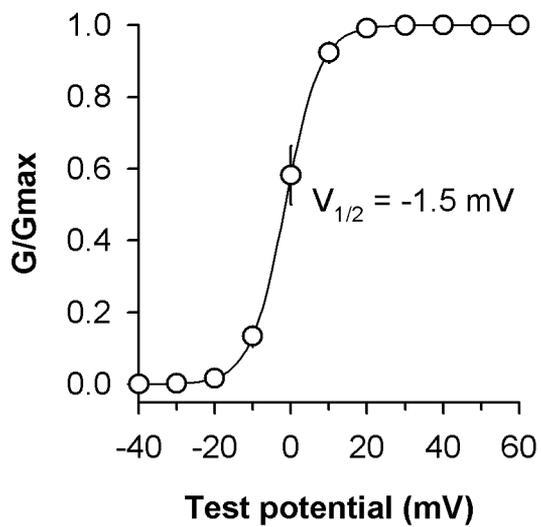

**E**

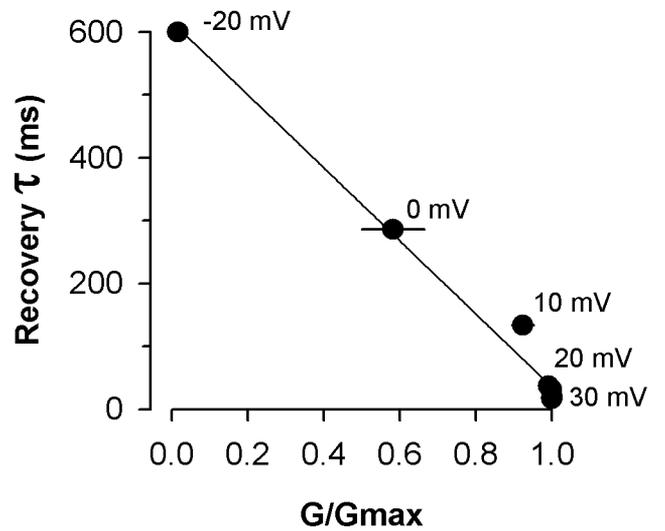

**A**

Ca$_V$2.2

30 mV

-90 mV

DAMGO

Control

200 nA

100 ms

Ca$_V$2.2 / β$_3$

DAMGO

Control

200 nA

100 ms

**B**

DAMGO inhibition (%)

**

(n=109)

(n=49)

Ca$_V$2.2          Ca$_V$2.2 / β$_3$

**C**

DAMGO inhibition (%)

***
***
***
***
***

***

**

NS

Time after start of depolarisation (ms)

**D**

DAMGO inhibition (%)

NS

(n=109)          (n=49)

Ca$_V$2.2          Ca$_V$2.2 / β$_3$

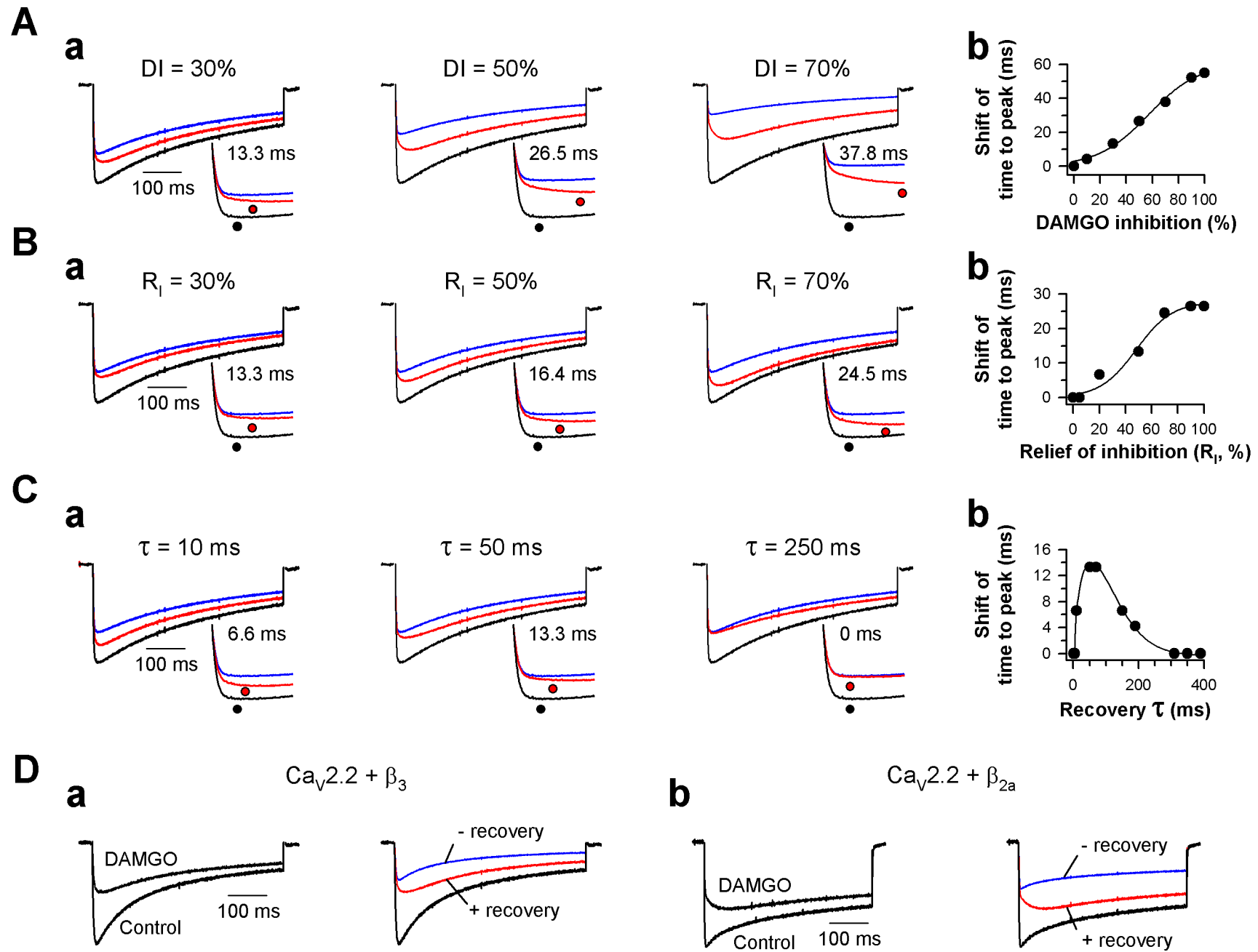

**A**

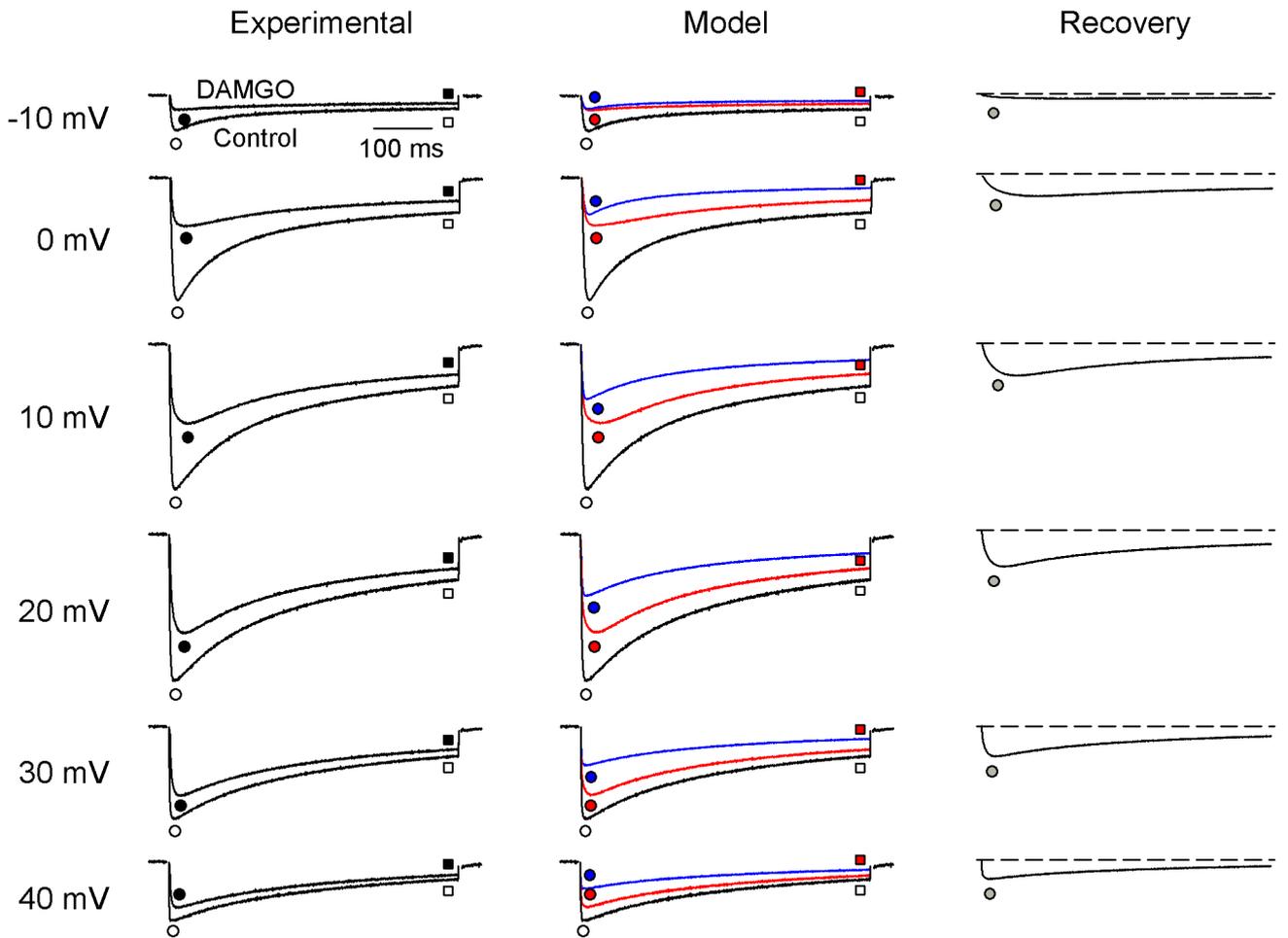

**B**

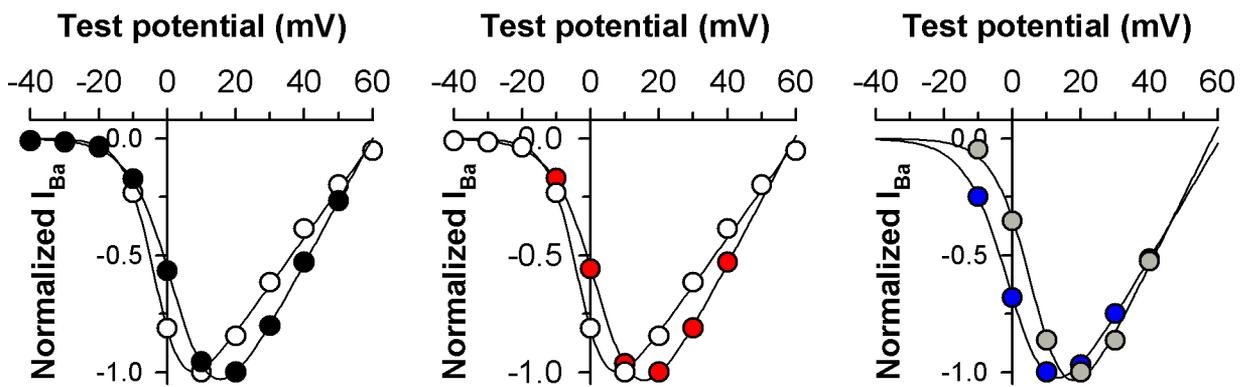

**C**

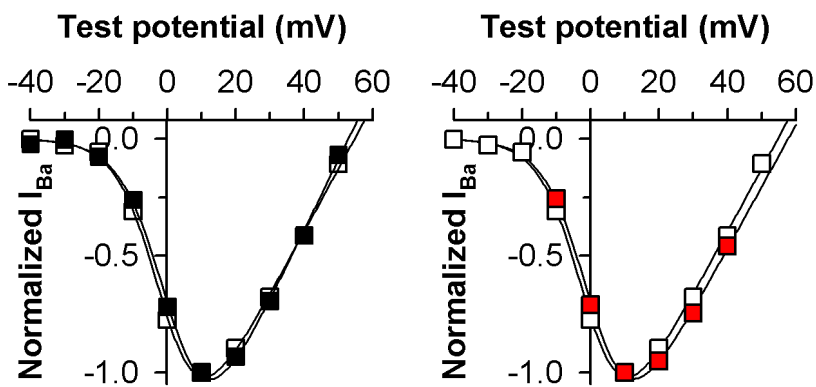

**A**

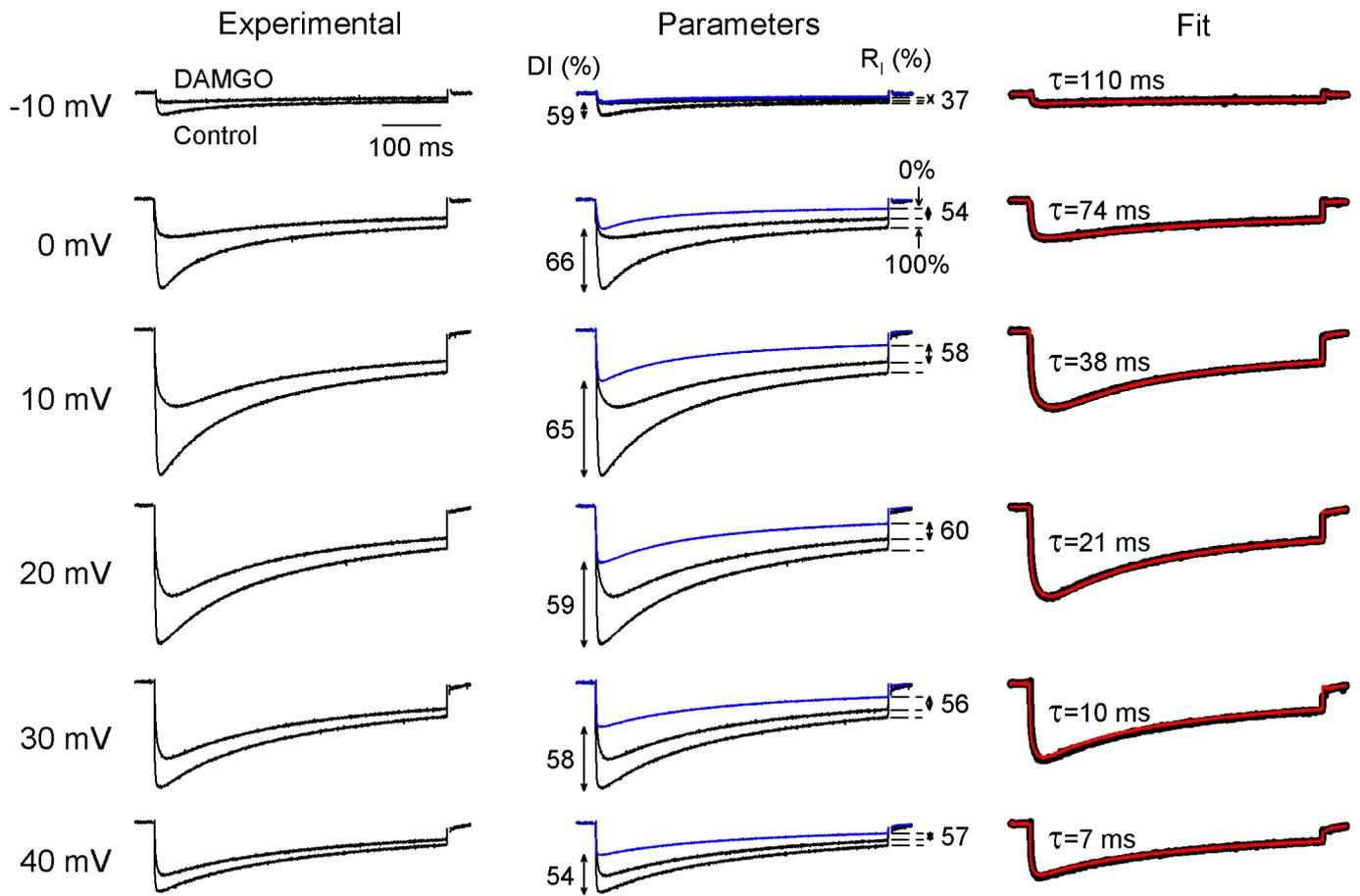

**B**

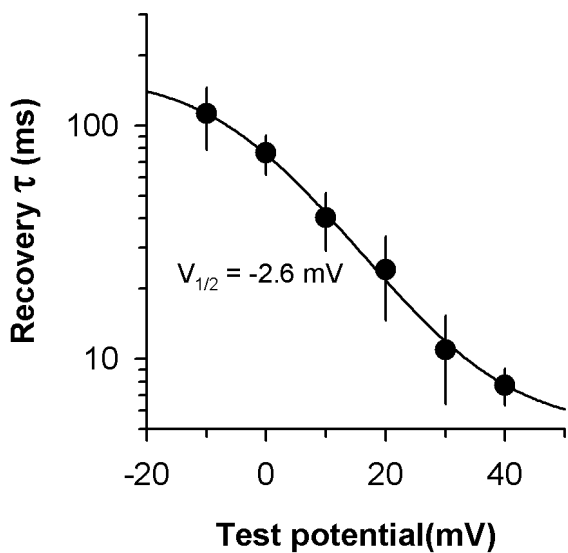

**C**

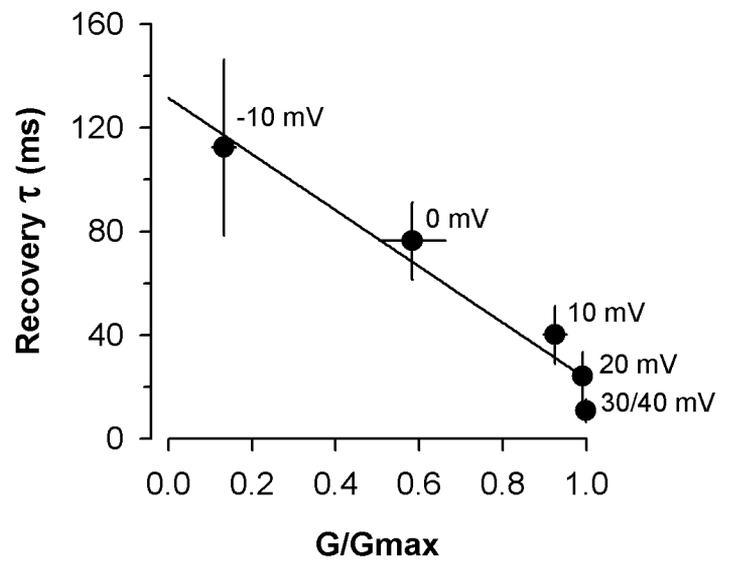

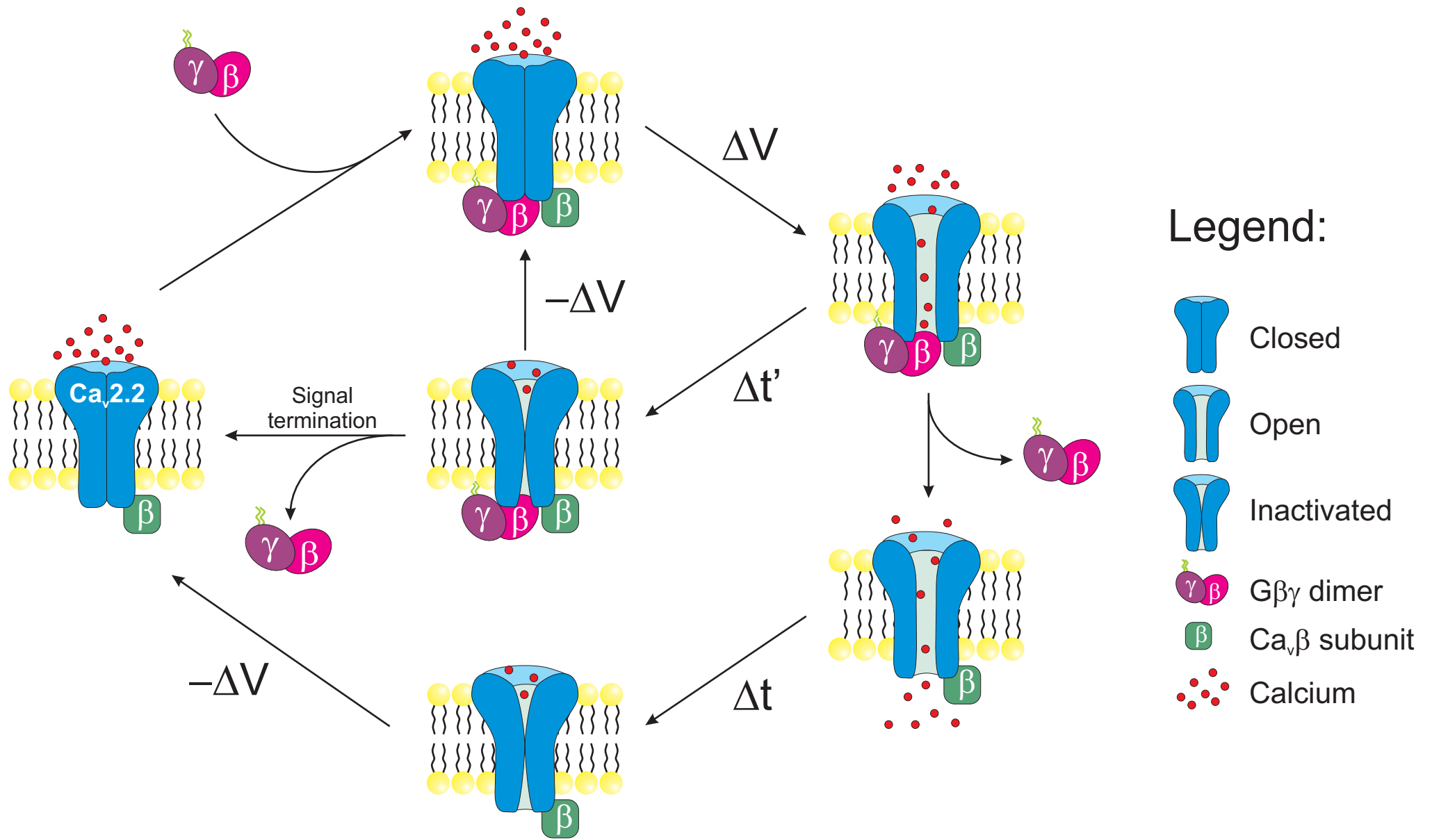

Legend:

Closed

Open

Inactivated

Gβγ dimer

Ca$_v$β subunit

Calcium